\documentclass[namedreferences]{kluwer}
\usepackage{epsfig,rotating}

\def\divB{\nabla\cdot\mathbf{B}}

\begin{document}
\begin{article}
\begin{opening}         

\title{Smoothed Particle Magnetohydrodynamics: \\ Some shocking results\ldots}
\author{D.J. \surname{Price}$^1$, J.J. \surname{Monaghan}$^2$}
\runningauthor{Price \& Monaghan}
\runningtitle{Smoothed Particle Magnetohydrodynamics}
\institute{$^1$Institute of Astronomy, Madingley Rd, Cambridge, CB3 0HA, UK \\
$^2$School of Mathematical Sciences, Monash University, Clayton 3800, Australia}

\begin{abstract}
 There have been some issues in the past in attempts to simulate magnetic fields
using the Smoothed Particle Hydrodynamics (SPH) method. SPH is well suited to
star formation problems because of its Lagrangian nature. We present new, stable
and conservative methods for magnetohydrodynamics (MHD) in SPH and present
numerical tests on both waves and shocks in one
dimension to show that it gives robust and accurate results.
\end{abstract}


\end{opening}           

\bibliographystyle{klunamed}

\section{Introduction}
 Smoothed Particle Hydrodynamics (SPH) is a unique method of solving the
equations of gas dynamics in that it involves no spatial grid. For this reason 
SPH is ideally suited to star formation type problems. However
there have been some issues in the past with attempts to simulate magnetic
fields in SPH, namely that when a conservative formulation of the momentum
equation was used an instability developed which causes particles to clump
together unphysically \cite{pm85}.

 We resolve this instability by adding a short range repulsive force to prevent particles from
clumping \cite{monaghan00}. In addition we formulate the equations of Smoothed Particle
Magnetohydrodynamics (SPMHD) using the continuum equations of \inlinecite{janhunen00} and
\inlinecite{dellar01} which are consistent even when the divergence of the magnetic field is 
non zero. Consequently, even though non zero $\divB$ may be produced 
during the simulation, it is treated consistently.

 Shocks are captured within SPMHD by use of artificial dissipation terms. The
formulation of these terms follows naturally by requiring dissipation in the
total energy across a shock front and then demanding that these terms result in
positive definite changes to the entropy. The artificial dissipation is
effectively turned off away from shocks by use of the switch proposed by \inlinecite{mm97}.

 The resulting equations, when implemented with a simple predictor 
corrector scheme for the timestepping, give good results for a wide range of shock tube 
problems \cite{pm03a,pm03b}.  While we have yet to apply our algorithm to problems in two 
and three dimensions the present results encourage us to believe that 
our SPMHD code will provide a secure basis for astrophysical MHD 
problems.

\section{Numerical tests in one dimension}
 The numerical scheme has been tested on a wide variety of one
dimensional problems. We present results here on three standard tests
which have been used to test many grid-based astrophysical MHD codes 
(e.g. \opencite{sea92}; \opencite{dw94}; \opencite{rj95}; \opencite{balsara98}).
For one dimensional MHD problems the magnetic field and velocity are allowed to
vary in three dimensions.
 
\begin{figure*}[t]
\begin{center}
\epsfig{file=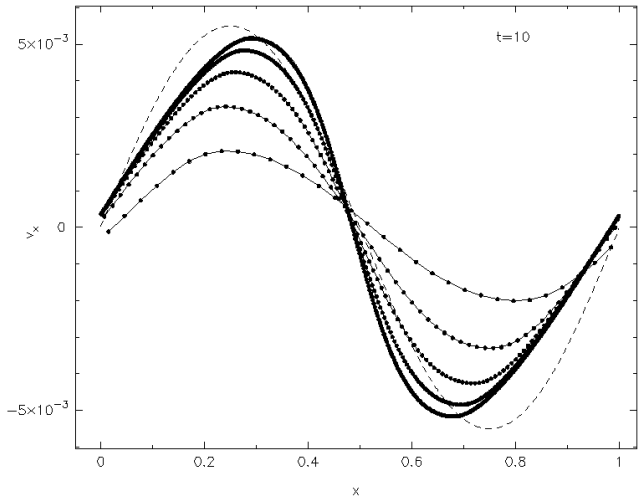,width=0.8\textwidth}
\caption{A travelling MHD fast wave propagating in a periodic 1D domain. Initial conditions
are indicated by the dashed line whilst the SPMHD solution is given by the solid
points after 10 periods (corresponding to 10 crossings of the domain), at resolutions of 32, 64, 128, 256 and 512
particles. The numerical solutions are in phase with the initial conditions,
demonstrating that the speed at which the wave travels is correctly captured by
our method. There is a small amount of steepening present due to non-linear effects,
which is in accordance with the solution presented in Dai and Woodward (1998).}
\label{fig:fastwave}
\end{center}
\end{figure*}

 The first test involves a fast MHD wave
propagating in a periodic 1D domain and is taken from \inlinecite{dw98}. The
wave is evolved for 10 periods, corresponding to 10 crossings of the
domain. The velocity profile in the
SPMHD solution is shown in Figure
\ref{fig:fastwave}, at resolutions of 32, 64, 128, 256 and 512 particles (solid points indicate the particles).
Note that the numerical solutions are in phase with the
initial conditions (dashed line), demonstrating that the speed at which the wave
travels is correctly captured by our method. This improvement over previous SPH
results (e.g. \opencite{mphd96}) is obtained by allowing the
smoothing length to vary with local particle density and deriving
self-consistently (via a Lagrangian variational principle)
the additional terms which should therefore be included in the SPH \cite{sh02} and
correspondingly the SPMHD \cite{pm03b} equations.

\begin{figure*}[t]
\begin{center}
\epsfig{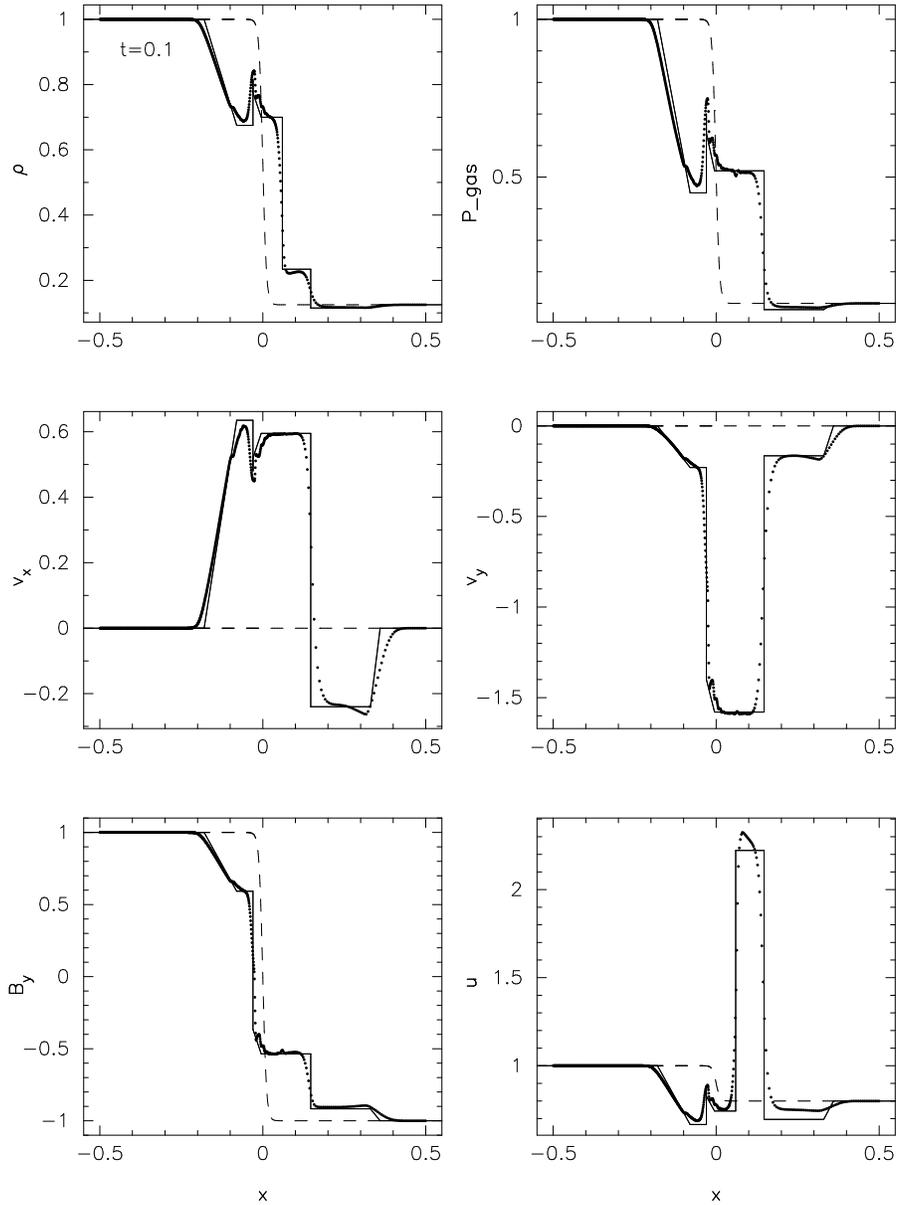}
\caption{Results of the Brio and Wu (1988) shock tube test. The problem consists of an
initial discontinuity at the origin, with two phases of gas brought into contact
at $t=0$. The initial conditions to the left of the shock are $(\rho,P,v_x,v_y,B_y) =
[1,1,0,0,1]$, whilst to the right the conditions are $(\rho,P,v_x,v_y,B_y)=[0.125,0.1,0,0,-1]$ with
$B_x = 0.75$ everywhere and $\gamma=2.0$. These initial conditions are shown by
the dashed lines, whilst the solid points, corresponding to the SPH particles, indicate the profiles of density, pressure, $v_x$, $v_y$,
transverse magnetic field $B_y$ and thermal energy $u$ at time $t=0.1$. The simulation uses 800
particles in the one dimensional domain x = [-0.5, 0.5]. The dissipation switch has not been
used in this case. The solution corresponds well to that computed using a
grid-based code incorporating a Riemann solver (Balsara 1998 - solid lines).}
\label{fig:briowu}
\end{center}
\end{figure*}

 The second test (Figure \ref{fig:briowu}) was first described by \inlinecite{bw88} and is the MHD analog of the
\inlinecite{sod78} shock tube problem. The problem consists of a
discontinuity in pressure, density, transverse magnetic field and
internal energy initially located at the origin. As time develops complex shock structures
develop which only occur in MHD because of the different wave types. The results
shown compare well with the solution computed by \inlinecite{balsara98} using a
grid-based MHD code incorporating a Riemann solver, indicated by the solid lines. 

\begin{figure*}[t]
\begin{center}
\epsfig{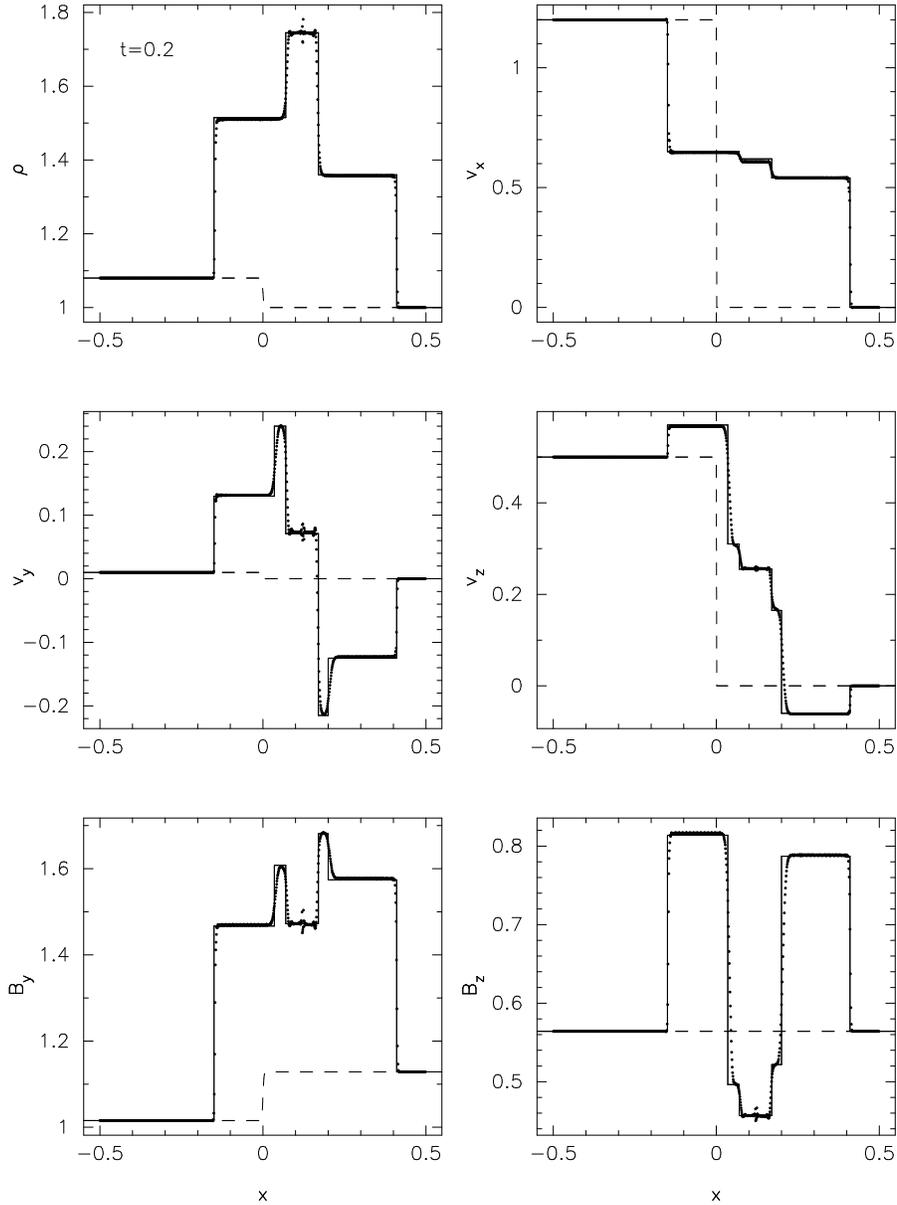}
\caption{Results of the isothermal MHD shock tube test. This problem illustrates the formation
of six discontinuities in isothermal MHD. The setup is similar
to that in Figure \ref{fig:briowu} with gas left of the origin initially in the state
$(\rho,v_x,v_y,v_z,B_y,B_z) =
[1.08,1.2,0.01,0.5,3.6/(4\pi)^{1/2},2/(4\pi)^{1/2}]$ and gas to the right
in the state
$(\rho,v_x,v_y,v_z,B_y,B_z)=[1,0,0,0,4/(4\pi)^{1/2},2/(4\pi)^{1/2}]$ with $B_x =
2/(4\pi)^{1/2}$ everywhere. These initial conditions are shown by the dashed lines,
whilst points indicate the positions of the SPH
particles at time $t=0.2$ which are shown to agree well with the numerical solution computed
by Balsara (1998) using a grid-based code (solid lines).}
\label{fig:balsara5}
\end{center}
\end{figure*}

 The third example is a test of the code for isothermal
MHD and is compared with the solution computed by \inlinecite{balsara98} in Figure \ref{fig:balsara5}. The problem illustrates the formation of
six discontinuities in the same problem. 

\section{Future work}
 We hope to apply the algorithm to a large simulation of star
formation similar to that performed by \inlinecite{bbb03}, including the effects
of magnetic fields. This should provide significant insight into the role that
magnetic fields play in the star formation process.

\section*{Acknowledgements} 
DJP acknowledges the support of the Association of Commonwealth Universities and the Cambridge
Commonwealth Trust. He is supported by a Commonwealth Scholarship and Fellowship
Plan. We also thank the referee, Enrique Vazquez-Semadeni, for useful
suggestions which have helped to improve this paper.

\bibliography{/home/dprice/bibtex/sph,/home/dprice/bibtex/mhd,/home/dprice/bibtex/starformation}

\end{article}
\end{document}